\documentclass{aa}  

\usepackage{graphicx}
\usepackage{txfonts}
\usepackage{xcolor}
\usepackage{hyperref}
\usepackage{soul}

\begin{document} 

\title{The scintillating tail of comet C/2020 F3 (Neowise)}

\author{R.A. Fallows \inst{1,3} \and
B.\,Forte \inst{2} \and
M.\,Mevius \inst{1} \and
M.\,A.\,Brentjens \inst{1} \and
C.\,G.\,Bassa \inst{1} \and
M.\,M.\,Bisi \inst{3} \and
A.\,Offringa \inst{1} \and
G.\,Shaifullah \inst{4,5,6} \and
C.\,Tiburzi \inst{6} \and
H.\,Vedantham \inst{1} \and
P.\,Zucca \inst{1}}

\institute{ASTRON - the Netherlands Institute for Radio Astronomy, Oude Hoogeveensedijk 4, 7991PD Dwingeloo, the Netherlands \\ \email{rafallows@gmail.com} \and
Department of Electronic and Electrical Engineering, University of Bath, Claverton Down, Bath, BA2 7AY, UK \and
RAL Space, United Kingdom Research and Innovation – Science \& Technology Facilities Council – Rutherford Appleton Laboratory, Harwell Campus, Oxfordshire, OX11 0QX, UK \and
Dipartimento di Fisica ``G. Occhialini'', Universit\`a di Milano-Bicocca, Piazza della Scienza 3, 20126 Milano, Italy \and
INFN, Sezione di Milano-Bicocca, Piazza della Scienza 3, I-20126 Milano, Italy \and
INAF, Osservatorio Astronomico di Cagliari, Via della Scienza 5, 09047 Selargius, Italy }

\date{Received May 31, 2022; accepted July 31, 2022}

\abstract
{The occultation of a radio source by the plasma tail of a comet can be used to probe structure and dynamics in the tail.  Such occultations are rare, and the occurrence of scintillation, caused by small-scale density variations in the tail, remains somewhat controversial.}
{We present a detailed observation taken with the Low-Frequency Array (LOFAR) of a serendipitous occultation of the compact radio source 3C196 by the plasma tail of comet C/2020 F3 (Neowise).  3C196 tracked almost perpendicularly behind the tail, providing a unique profile cut only a short distance downstream from the cometary nucleus itself.}
{Interplanetary scintillation (IPS) is observed as the rapid variation of the intensity received of a compact radio source due to density variations in the solar wind. We observed IPS in the signal received from 3C196 for five hours, covering the full transit behind the plasma tail of comet C/2020 F3 (Neowise) on 16 July 2020, and allowing an assessment of the solar wind in which the comet and its tail are embedded.}
{The results reveal a sudden and strong enhancement in scintillation which is unequivocally attributable to the plasma tail.  The strongest scintillation is associated with the tail boundaries, weaker scintillation is seen within the tail, and previously unreported periodic variations in scintillation are noted, possibly associated with individual filaments of plasma.  Furthermore, contributions from the solar wind and comet tail are separated to measure a sharp decrease in the velocity of material within the tail, suggesting a steep velocity shear resulting in strong turbulence along the tail boundary.}
{}


\maketitle
%

\section{Introduction}

Interplanetary scintillation (IPS - \citet{Clarke:1964,Hewishetal:1964}), observed as the rapid variation of intensity received from a compact radio source due to density variations in the solar wind, has been used for several decades to observe the solar wind throughout the inner heliosphere.  Such observations are typically used to measure solar wind velocity \citep[e.g.][]{Coles:1996,ManoAnanth:1990,KojimaKakinuma:1990}, or to use the strength of the scintillation as a proxy for solar wind density \citep[e.g.][]{Jacksonetal:1998,Tappin:1986}.  The occultation of a radio source by the plasma tail of a comet provides an opportunity to use this phenomenon to probe turbulent-scale density structure and dynamics in the tail.  However, such occultations are rare, and the occurrence of scintillation attributable directly to the tail itself remains somewhat controversial.  

Positive results, where scintillation attributable to a comet plasma tail could be seen, were reported for comets Kohoutek \citep{Ananthakrishnanetal:1975}, Halley \citep{Alurkaretal:1986,Sleeetal:1987}, Wilson \citep{Sleeetal:1990}, Austin \citep{Janardhanetal:1991, Prasad:1994}, Schwassmann-Wachmann 3-B \citep{Royetal:2007}, and ISON (C/2012 S1) \citep{Ijuetal:2015}.  However, negative detections have also been reported for comet Halley by \citet{Ananthakrishnanetal:1987} and there was some debate over the positive results for this comet expressed in the correspondence section of Nature \citep{Ananthakrishnanetal:1989}.  Furthermore, \citet{HajivassiliouDuffett-Smith:1987} questioned all of the results published prior to their paper, claiming various inconsistencies that led them to conclude that there was no convincing evidence for enhanced scintillation due to comet ion tails.  \citet{Sleeetal:1990} revisited their earlier results and those of \citet{HajivassiliouDuffett-Smith:1987} and were able to account for the majority of the negative detections reported.  All of these results, and the debate surrounding them, illustrate the importance of being able to unambiguously identify scintillation from the comet tail, and separate it from contributions from the solar wind in which it sits and, at lower observing frequencies, the ionosphere.

An opportunity arose on 16 July 2020 to observe an occultation of the strong radio source 3C196 by the tail of comet C/2020 F3 (Neowise) using the Low-Frequency Array (LOFAR - \citet{vanHaarlemetal:2013}.  As an instrument with a large number of stations across western Europe and capable of recording wideband high-resolution dynamic spectra from each station individually, LOFAR is ideally suited to observing interplanetary scintillation \citep{Fallowsetal:2022a} and ionospheric scintillation \citep{Fallowsetal:2020}, with ways of separating the two \citep{Fallowsetal:2016} where necessary; there also exists the possibility to use the wealth of such data already taken to identify rarer phenomena.  At this time, the nucleus of the comet passed at a closest angular separation to the radio source of only $0\fdg42$.  Furthermore, the track of 3C196 relative to the comet tail was close to being perpendicular to the tail, meaning that the result is an almost direct cut across the tail.  This is in marked contrast to almost all prior results, where the radio source tracks have tended to traverse more along the tail. 

The LOFAR observation had a duration of five hours, allowing time both before and after the occultation for an assessment of background solar wind conditions, and enabling the additional contribution to the scintillation pattern from the comet tail to be obvious in the observation.  The result is therefore a unique observation which represents a natural laboratory for scattering theory, with the results having implications for the wider interpretation of radio scattering across both interplanetary and ionospheric scintillation regimes.

\section{Observation and initial processing}
\label{sec:observation}

\begin{figure}
    \centering
    \includegraphics[width=\linewidth]{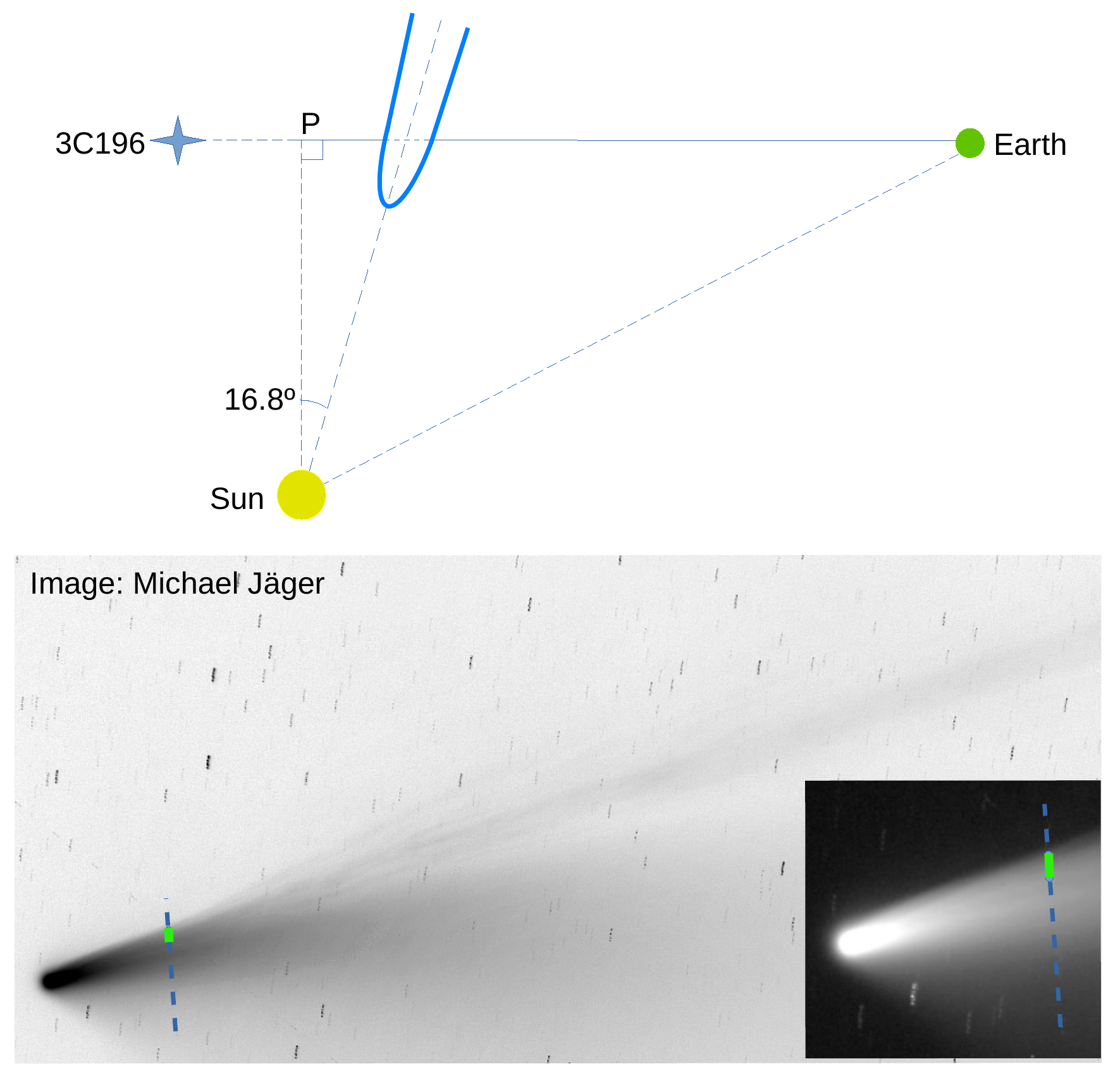}
    \caption{Top: Geometry of the line of sight to 3C196 relative to the Sun and comet (not to scale). P represents the point of closest approach of the line of sight to the Sun. Bottom: Track of 3C196 relative to the comet (blue dashed line), in the sky plane, overplotted on an image of the comet taken at 20:46\,UT on 17 July 2020 by Michael J\"ager (reproduced with permission). The green segment of the track marks the period of enhanced scintillation detailed in Section \ref{sec:results}. }
    \label{fig:diagram}
\end{figure}

3C196 was observed from 15:30\,UT to 19:14\,UT on 16 July 2020, covering the full transit behind the plasma tail of comet C/2020 F3 (Neowise).  At this time, 3C196 was at an elongation of $27\fdg2$ from the Sun and the comet passed at a closest angular separation to the radio source of only $0\fdg42$. Figure \ref{fig:diagram} illustrates the geometry in the Sun--Earth--line-of-sight plane, and plots the track of 3C196 relative to the comet in the sky on a photograph taken of the comet on the following evening (Image credit: Michael J\"ager, reproduced with permission).

The proposed observation setup was designed to achieve two objectives: one was to investigate whether or not any refraction of the radio source due to the plasma tail could be directly detected from an interferometric analysis, and the other was to assess the intensity scintillation pattern received for any contribution that could be attributable to it.  The first objective required interferometric data be collected and it was hoped that beamformed data could be taken simultaneously for the second.  Unfortunately, this proved beyond the limits of the system at that time.  The LOFAR data were therefore taken using only the interferometry mode, recording visibilities (correlated complex voltages from X- and Y- polarisations) for all stations (core stations, remote stations with the exception of RS306 and RS509, and international stations with the exception of PL611) available for observation at the time, with a time integration of 0.167\,s over a bandwidth of 78.125\,MHz between 110 and 189\,MHz, divided into frequency bins of 12.2\,kHz.  In LOFAR observation-specification parlance, this is 400 sub-bands (each 195.3125\,kHz wide) subdivided into 16 channels per sub-band.  For the purposes of analysing the data for intensity scintillation, the visibilities from the auto-correlations were converted into Stokes intensity via a standard formula, $I = re(XX+YY),$ and averaged to one channel per sub-band by taking the tenth percentile of the intensities.  This has the effect of reducing contributions from radio frequency interference (RFI) and the resulting intensity dynamic spectra had a final time resolution of 0.167\,s and frequency resolution of 195.3125\,kHz, and were saved to a separate HDF5 file.  An intensity dynamic spectrum was also generated via the same method for visibilities formed from the coherent sum of all core stations (forming a `tied-array beam' in LOFAR parlance).  The raw and processed visibilities and HDF5 dynamic spectra are stored in the LOFAR long-term archive\footnote{\url{https://lta.lofar.eu/Lofar}} under project code DDT14\_001, where they are available for download.  Whilst it is recognised that this time resolution is not ideal for interplanetary scintillation, it proved sufficient in this instance for the analyses described here.

Processing of the intensity dynamic spectra to produce a set of time series for power spectra calculation and cross-correlation follows the same basic methods described in \citet{Fallowsetal:2022a} and \citet{Fallowsetal:2020}, with some minor differences.  RFI remaining in the dynamic spectra was identified using the following steps:- 

\begin{itemize}
    \item Applying a median filter with a [time,frequency] dimension of [3.34\,s,1.95\,MHz], corresponding to [20,10] data points;
    \item Subtract the median filtered data from the original to form a flattened array, normalised to 1.0;
    \item Calculate the standard deviation of the flattened data about the median;
    \item Apply a threshold of 20 standard deviations;
    \item Frequency channels where the mean of normalised RFI-excluded points is less than 0.8 are flagged in their entirety.
\end{itemize}

Whilst the threshold might seem quite high, it should be noted that the standard deviation about the median is a small value and the threshold must be low enough to exclude the majority of RFI but high enough not to falsely identify strong peaks in the scintillation intensity pattern.  The threshold multiple of 20 was identified as being reasonable based on trial and error.  The original data, with identified RFI now flagged were detrended in time by dividing each frequency channel by a third-order polynomial fitted to the channel data.

The scintillation pattern received at Earth has a drift velocity resulting from the solar wind flowing in a direction perpendicular to the lines of sight between radio source and receivers. This drift velocity therefore represents a line-of-sight integration of solar wind velocity components perpendicular to the lines of sight.  However, as the amount of scattering falls with distance from the Sun as $R^{-4}$, it is commonly assumed that the drift velocity corresponds closely to the solar wind velocity around the point of closest approach of the line of sight to the Sun (the P-point labelled in \ref{fig:diagram}), and this assumption is broadly reasonable if a single solar wind stream is dominant in the line of sight.  A cross-correlation analysis represents the most accurate method to calculate the velocity, as detailed in \citet{Fallowsetal:2022a} and references therein.

For the cross-correlation analysis, intensity time series were calculated by taking the median of the dynamic spectra across the 150-169\,MHz section of the band, chosen such that the intensity structure remains well correlated over the band and strong interference from digital broadcast signals at the higher frequencies is excluded.  A subset of predominantly international stations (Core, RS508, DE602, DE603, DE604, DE605, FR606, SE607, UK608, IE613, and LV614) were used in this analysis; remaining stations were found to have a higher level of interference which adversely affected the analysis.  The intensity time series were divided into five-minute segments, advanced every 10\,s.  A high-pass filter at 0.05\,Hz (this choice is lower than normally applied to IPS data, but is justified by the power spectra presented in subsection \ref{subsec:pspec}) and low-pass filter at 2.3\,Hz were applied to the power spectra calculated from the time series before cross-correlation.

\section{Results}
\label{sec:results}

\subsection{Dynamic spectra}
\label{subsec:dynspec}

\begin{figure}
    \centering
    \includegraphics[width=\linewidth]{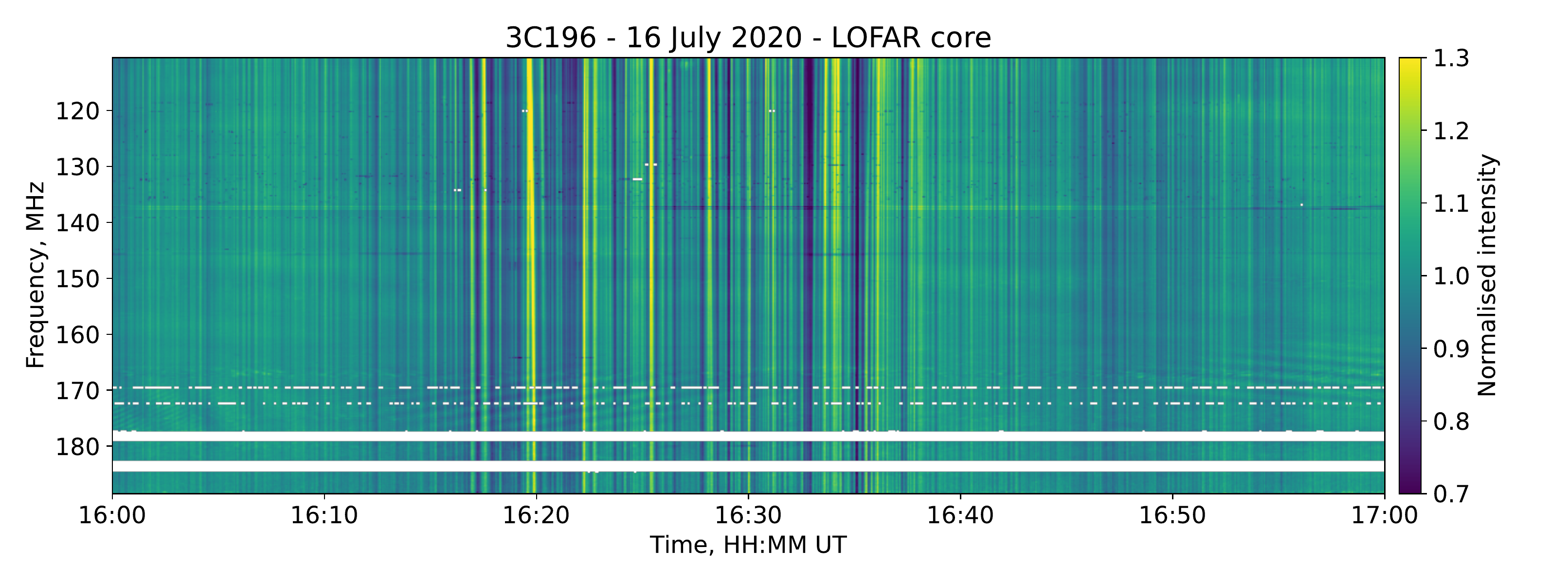}
    \includegraphics[width=0.49\linewidth]{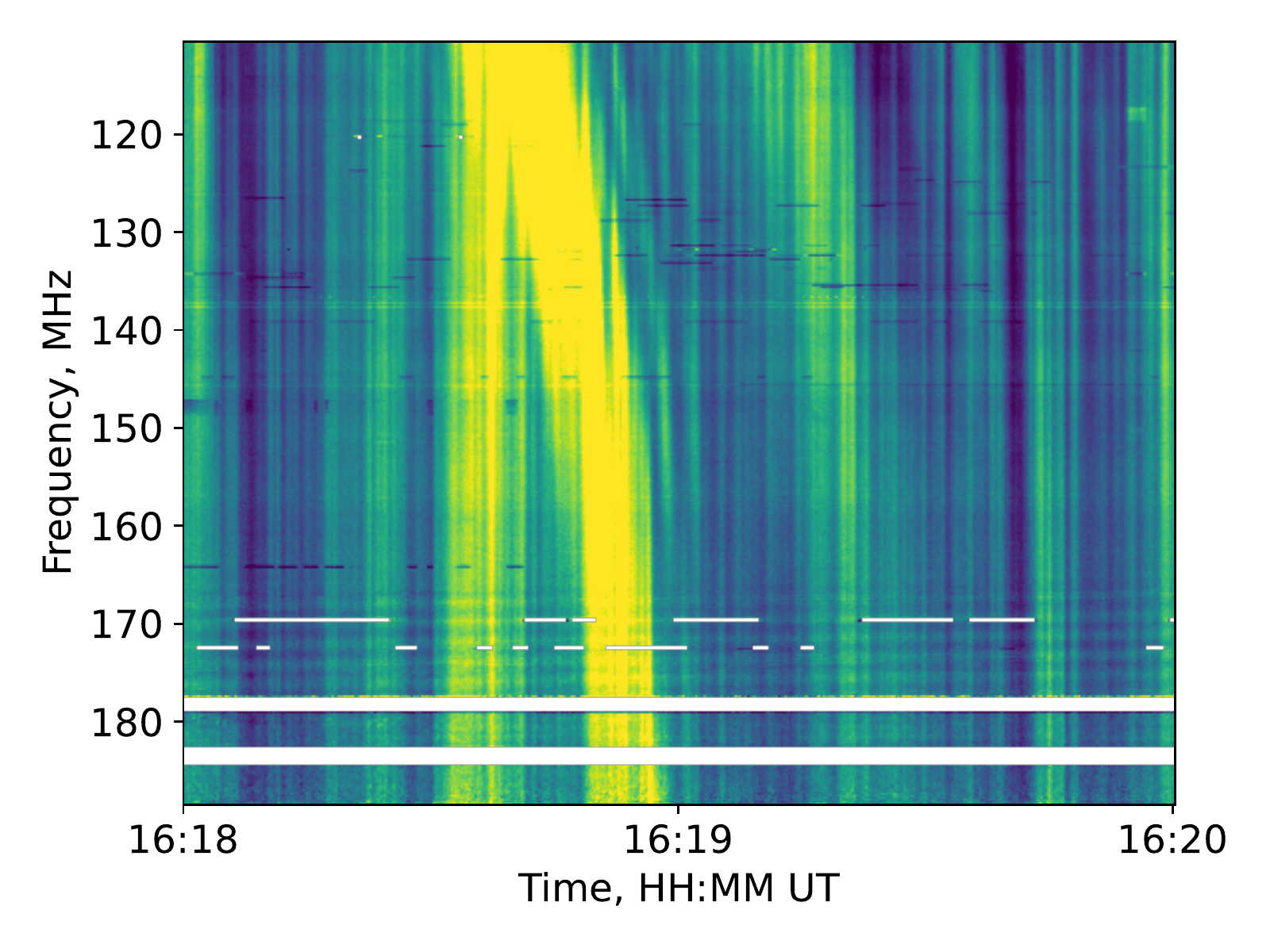}\includegraphics[width=0.49\linewidth]{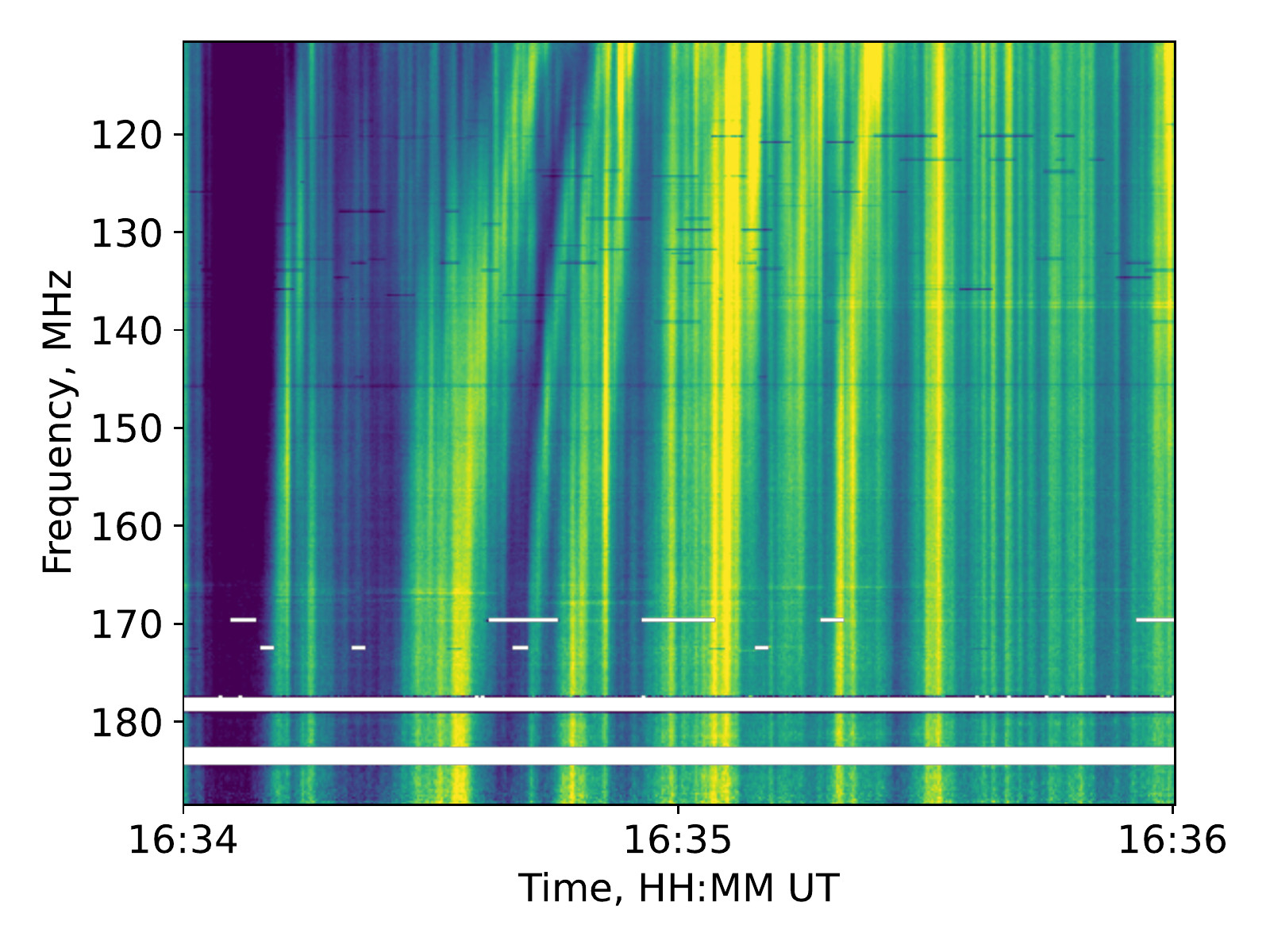}
    \caption{Dynamic spectra of the variation in normalised intensity received by the LOFAR core. Top: One hour of data showing the obvious enhancement in scintillation between approximately 16:15\,UT and 16:40\,UT. Lower left: Two minutes of data from near the start of the period of enhancement.  Lower right: Two minutes of data from near the end of the period of enhancement.}
    \label{fig:dynspec}
\end{figure}

Figure \ref{fig:dynspec} shows a dynamic spectrum of the variation in intensity received by the LOFAR core, along with details of particular features.  A period of strong scintillation lasting nearly 25 minutes is obvious in the figure.  The same pattern is seen in data from all international stations of LOFAR, covering an area from Ireland to Latvia, which immediately discounts the possibility of the ionosphere being the source of the enhanced scintillation.  Furthermore, no such short-duration burst of scintillation has been observed in many years of IPS data taken using LOFAR and other instruments by the primary author, including during extended observations covering the passage of coronal mass ejections. Therefore, the plasma tail of the comet is the most likely source of the scintillation, and this is further indicated by its occurrence corresponding to the period when 3C196 was immediately downstream from the comet, as illustrated in Figure \ref{fig:diagram}.

The majority of the scintillation is visible as a rapid variation that remains well-correlated across the observing band, but a change is obvious near the start and end of the period of enhancement (lower plots of Figure \ref{fig:dynspec}).  Here, the intensity shows a structure of  longer duration, curved towards lower frequencies, with opposite curvatures of the earlier and later times, demonstrating that intensity enhancements at the lower frequencies were first to arrive and last to leave.  Similar effects are seen weakly within the period of enhancement. 

\subsection{Scintillation indices}
\label{subsec:si}

\begin{figure}
    \centering
    \includegraphics[width=\linewidth]{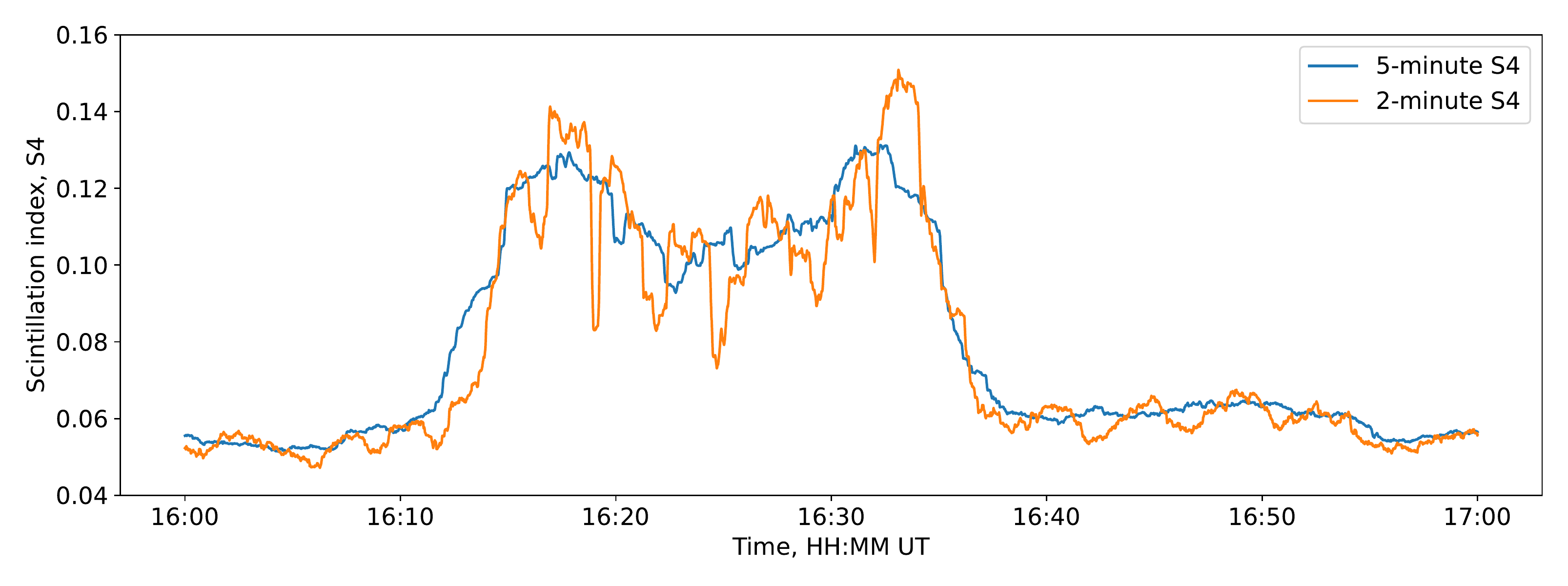}
    \caption{Scintillation indices calculated for segments of intensity data for the tied-array beam formed from the core stations averaged over the frequency band 150-190\,MHz, of both 2 min and 5 min duration.}
    \label{fig:s4}
\end{figure}

The scintillation index, a normalised measure of the strength of the scintillation, can be defined as \citep[][]{BriggsParkin:1963}:

\begin{equation}
    S_{4}^{2} = \frac{\langle I^{2} \rangle - \langle I \rangle^{2}}{\langle I \rangle^{2}}
,\end{equation}

\noindent where $I$ represents the zero-mean normalised intensity fluctuations; $\langle\rangle$ denotes ensemble averaging, which is substituted by temporal averaging in the case of experimental observations. Here, $S_{4}$ was estimated on the basis of the standard deviation of the zero-mean normalised intensity fluctuations over time intervals of both 5 min and 2 min for all the radio wave frequencies observed, using the data from the tied-array beam formed from the core stations.  Figure \ref{fig:s4} shows a median of indices for frequencies above 150\,MHz.

An enhancement in scintillation is obvious between 16:12 and 16:37\,UT, showing a sharp twin-peak structure with a dip in the centre.  A periodic structure is seen in the two-minute indices, which is somewhat averaged out in the five-minute calculations.  This period of enhancement corresponds to the green segment of track of 3C196 relative to the comet given in Figure \ref{fig:diagram}, illustrating a strong correspondence with the passage of the plasma tail over the line of sight.

\begin{figure}
    \centering
    \includegraphics[width=\linewidth]{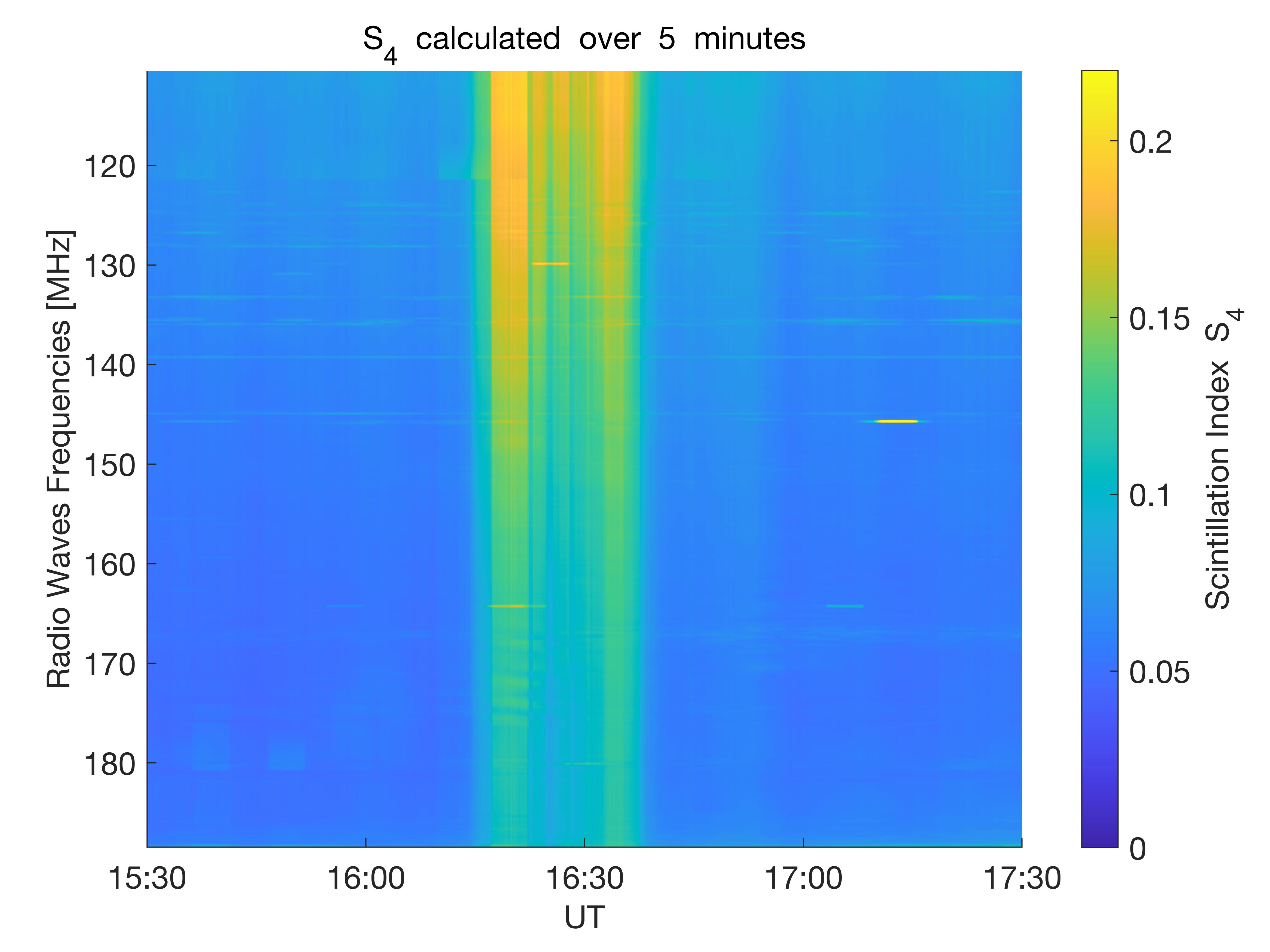}
    \includegraphics[width=\linewidth]{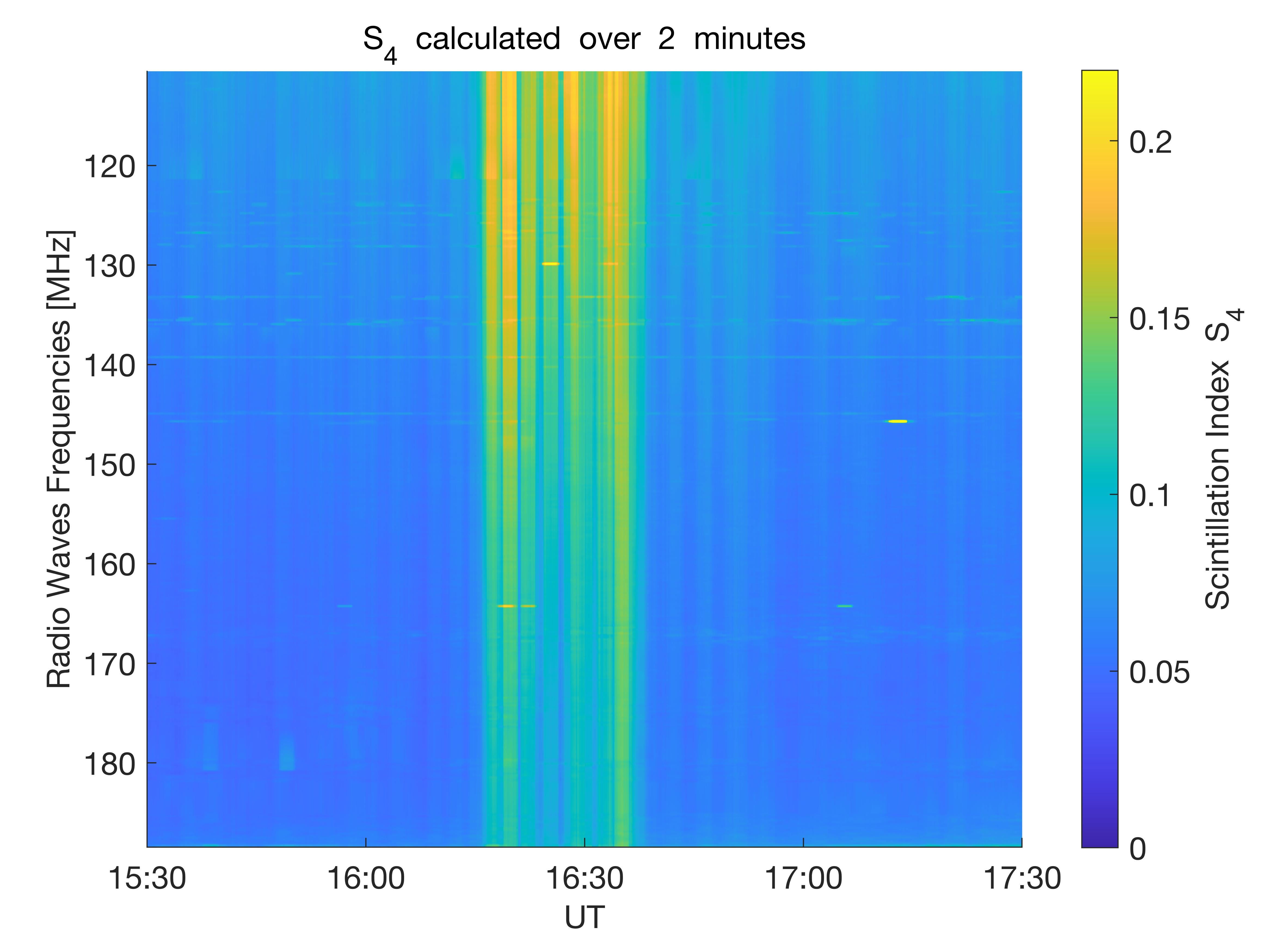}
    \caption{ $S_{4}$ scintillation index estimated for scintillation occurring on radio wave frequencies when intersecting comet C/2020 F3 (NEOWISE) and observed by the tied-array beam formed from the core stations between 15:30 and 17:30\,UT on 16 July 2020.  Top: S4 calculated over  intervals of 5 min.  Bottom: S4 calculated over  intervals of 2 min.}
    \label{fig:s4_3_new}
\end{figure}

This is further illustrated in Figure \ref{fig:s4_3_new}, which shows the S4 indices for all frequencies observed.  The enhancement in the scintillation index $S_{4}$ is characterised across the observing band by the two main peaks, which correspond to the intersection of the line of sight with the outer edges of the plasma tail.  However, towards the lower frequencies, the scintillation indices become stronger (as would be expected within the weak scattering regime) and the dip in between the two main peaks becomes less distinct.  Indeed the multiple peaks of the two-minute indices almost reach the same maximum as the main peaks at the lowest frequencies. 



\subsection{Power spectra}
\label{subsec:pspec}

\begin{figure}
    \centering
    \includegraphics[width=0.49\linewidth]{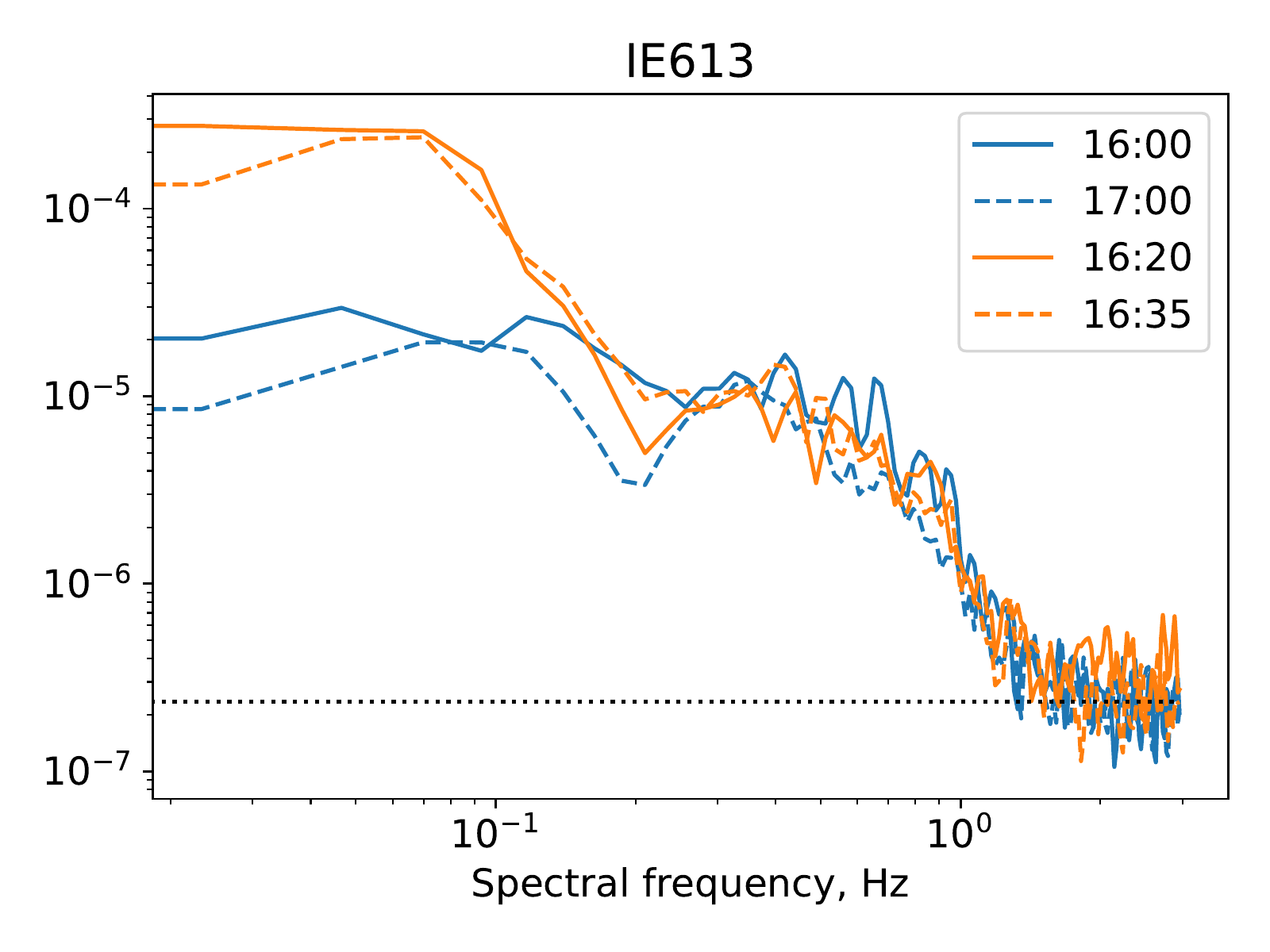}\includegraphics[width=0.49\linewidth]{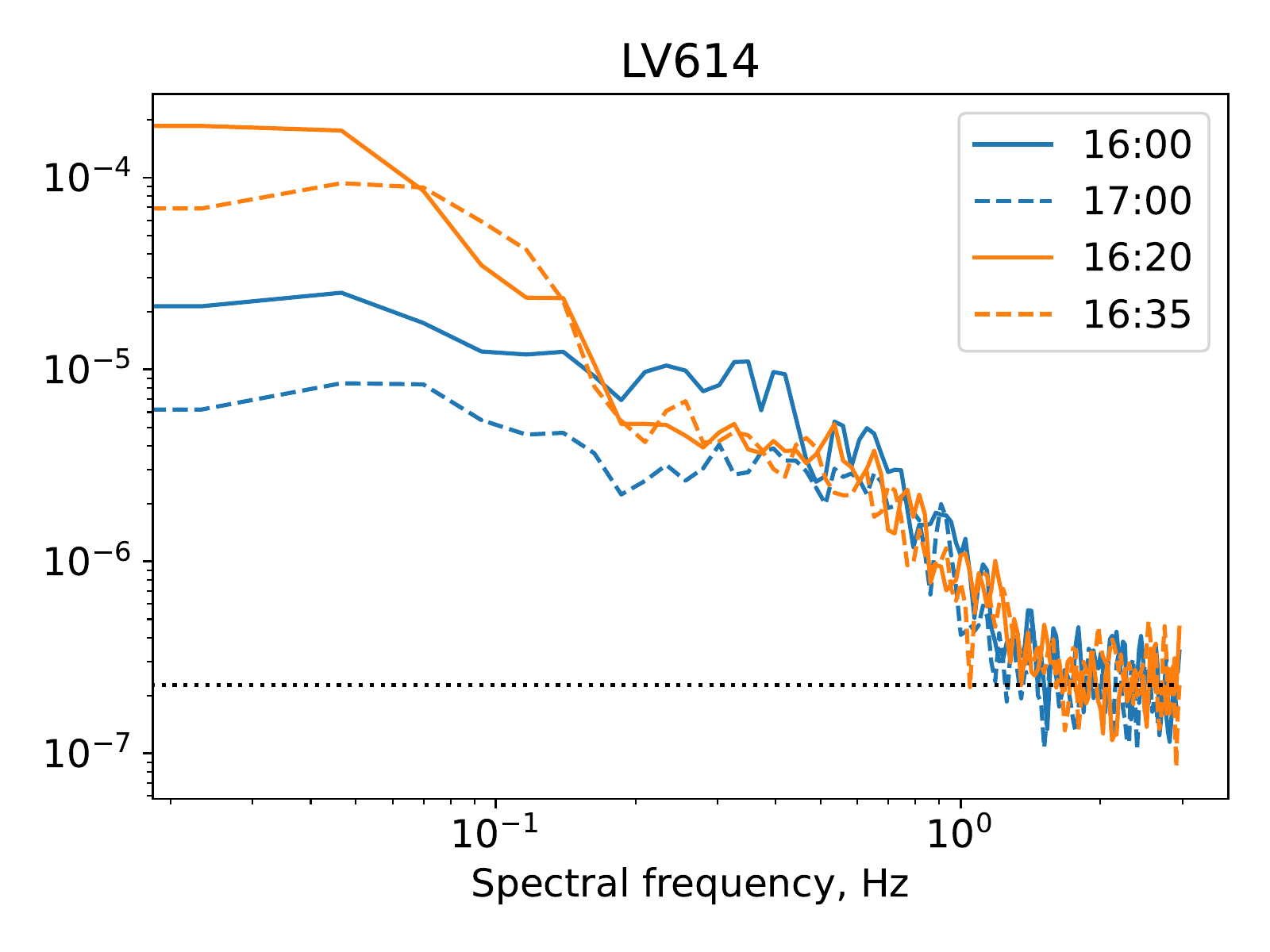}
    \caption{Power spectra from IE613 (left) and LV614 (right) for time intervals starting at 16:00\,UT, 16:20\,UT, 16:35\,UT, and 17:00\,UT, for five-minute segments of data averaged over the frequency band 150-169\,MHz.  The white noise floor is indicated by the black dotted lines for reference.}
    \label{fig:pspec}
\end{figure}

Figure \ref{fig:pspec} shows example power spectra for IE613 (Birr, Ireland) and LV614 (Ventspils, Latvia) for four time intervals, corresponding to times well before and after the period of enhanced scintillation and of the peaks in enhanced scintillation.

The scintillation power below 0.2\,Hz increases strongly in the spectra at 16:20\,UT and 16:35\,UT compared to times before and after for both stations.  Such an increase would normally be assumed to be ionospheric in nature \citep[e.g.][]{Fallowsetal:2016} but in this case it is present in the power spectra from all stations simultaneously and only during the period of enhanced scintillation.  Such a phenomenon could not be ionospheric, and so is assumed to be associated with the comet plasma tail for the purposes of the cross-correlation analysis.  The location of the high-pass filter was therefore moved from a more usual 0.2\,Hz to 0.05\,Hz for the cross-correlation analysis.

Given the low level of scintillation observed outside of the period of enhancement, the radio-wave propagation problem can be approximated during the period of enhancement by assuming a single phase-changing screen in the comet tail.  In this approach, fluctuations in the refractive index are due to fluctuations in the spatial distribution of the electron density.  A turbulent distribution of irregularities described through a power-law spatial spectrum within the phase screen then gives rise to asymptotic behaviour in the power spectrum of the intensity fluctuations.  Thus, the power spectrum of intensity fluctuations under the assumption of weak scatter typically exhibits a power-law similar to that of the irregularity spectrum for temporal frequencies greater than the Fresnel frequency (Equation \ref{eqn:Fresnel}), but approaches, approximately, a constant for temporal frequencies smaller than the Fresnel frequency.  This results in a characteristic knee in the spectrum at the Fresnel frequency, given by:

\begin{equation}
    \nu_{F} = \frac{V^{REL}}{\sqrt{2 \lambda z}} 
    \label{eqn:Fresnel}
,\end{equation}

where $V^{REL}$ is the relative drift between the ray path and the phase screen, $\lambda$ is the wavelength of the radio wave, and $z$ is the distance between the phase screen and the receiver.

By taking the wavelength as 2\,m (150\,MHz), the distances from Earth to the phase-changing screens as being 133352260\,km (191.6\,R$_{Sun}$) to the P-point of the line of sight and 112499000\,km as the distance to the comet at 16:25\,UT on 16 July 2020, and $\nu_{F}$ as 0.8\,Hz and 0.08\,Hz for the solar wind and comet plasma tail, respectively, a calculation of V$^{REL}$ yields 584\,km\,s$^{-1}$ for the solar wind and 54\,km\,s$^{-1}$ for the plasma tail of the comet.  It should be noted that these calculations become 1012\,m\,s$^{-1}$ and 101\,m\,s$^{-1}$ respectively if an ionospheric distance of 400\,km is assumed.  Whilst the former is unlikely to be taken as being reasonable for the mid-latitude ionosphere, the latter could be, illustrating a case of possible confusion \citep[detailed theoretically in][]{Forteetal:2022} between scintillation from different possible origins.


\subsection{Velocity}
\label{subsec:velocity}

\begin{figure}
    \centering
    \includegraphics[width=\linewidth]{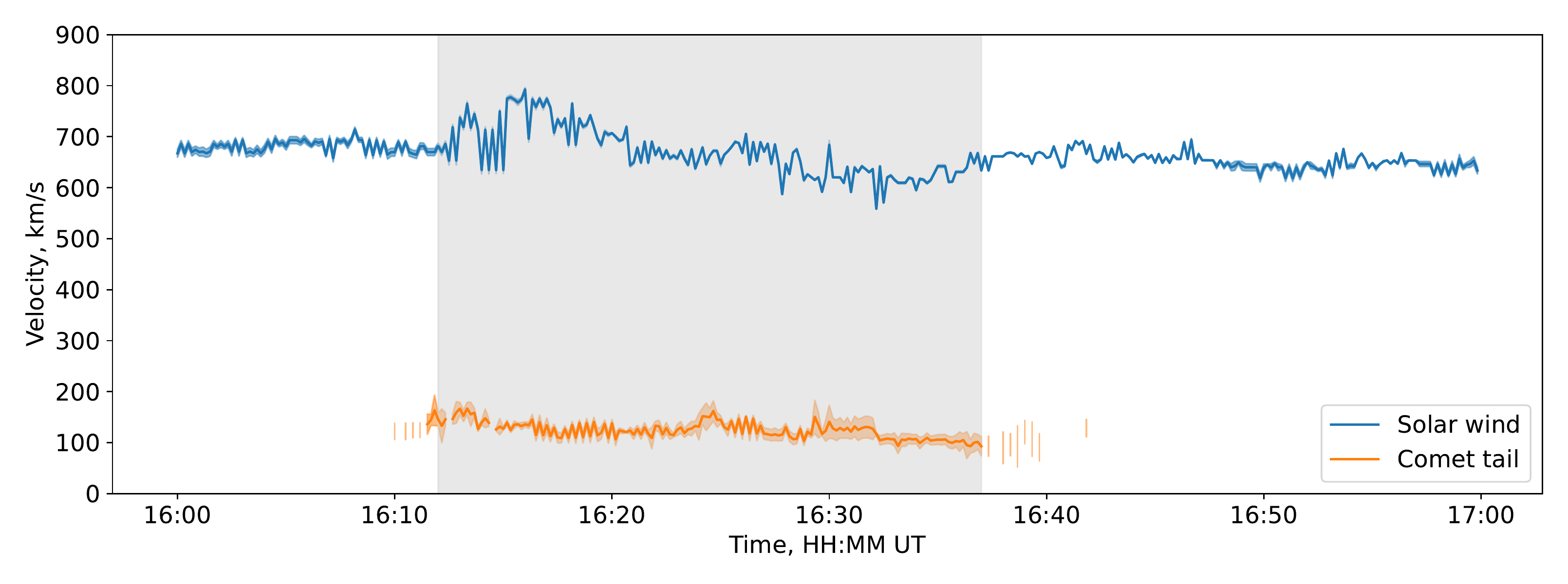}
    \caption{Velocity calculated from least-squares fits to plots of radial baseline versus cross-correlation peak time lag.  The velocity calculated for the background solar wind is plotted in blue, and that for the comet tail in orange.  The range in velocity in each case is the standard deviation calculated from the fits, although this is too small to be easily visible here.  The grey shaded area represents the period of enhanced scintillation.}
    \label{fig:velocities}
\end{figure}

Components of the baseline length between each pair of stations are calculated in terms of the radial direction from the Sun, and the direction perpendicular to it, for the baselines projected onto the sky plane.  As the solar wind is commonly assumed to be radial in direction from the Sun, least-squares fits to plots of the radial baseline versus the time lag of the peaks of the cross-correlation functions were performed to estimate velocity for each time segment through the observation, following the methods detailed in \citet{Fallowsetal:2022a}.  For the bulk of the observation, these plots show little scatter after obvious outliers ---typically  due to segments of poor data from one or more stations--- were excluded, demonstrating a single solar wind velocity.  However, during the period of enhanced scintillation, significant extra scatter is seen, indicating the presence of a different velocity.  Separate fits were therefore performed, excluding points consistent with the background solar wind velocity and any obvious further outliers.  The results from fits to both velocities are shown in Figure \ref{fig:velocities}. 

A main velocity of around 700\,km\,s$^{-1}$, consistent with a fast solar wind stream, is seen throughout the observation, with some variation during the period of enhanced scintillation.  A second velocity at a little over 100\,km\,s$^{-1}$ is found only during the period of enhanced scintillation, and was found to be prominent throughout this period only when the excess power below 0.2\,Hz in the spectra was included in the cross-correlation analysis.  If the high-pass filter was set at a more usual 0.2\,Hz this velocity was less obvious, particularly in the middle of the period of enhanced scintillation.

\subsection{Intensity interferometry}
\label{subsec:intensityinterferometry}

The primary purpose of interferometry in radio astronomy is to obtain a detailed image of the radio source or field being observed, which entails careful calibration for signal delays resulting from variations in the observing system itself and the effects of propagation through intervening media.  \citet{Fallowsetal:2022a} demonstrated that applying the techniques of intensity interferometry (involving the cross-correlation of signal intensities instead of complex voltages) can result in the visualisation of properties related to the turbulent structure giving rise to scintillation itself.  Full details of how this technique is used and applied are given in that paper and will not be repeated here, but we provide a brief summary below.  

For any time lag in the intensity, cross-correlation functions calculated as above the corresponding cross-correlation values can be plotted on a spatial grid with components in the radial direction from the Sun and tangential to it.  Each value is plotted on this grid according to the (radial,tangential) components of the baseline between the associated pair of stations.  Cross-correlation values at the positive time lag are plotted assuming the baseline vector is taken in the positive radial direction (i.e. away from the Sun), and the cross-correlation values at the equivalent negative time lag are simultaneously plotted assuming the baseline vector is in the negative radial direction (i.e. towards the Sun).  The resulting scatter plot is then contoured and the contours displayed as the final image, which is henceforth referred to as a `spatial correlation' image. 

\begin{figure}
    \centering
    \includegraphics[width=\linewidth]{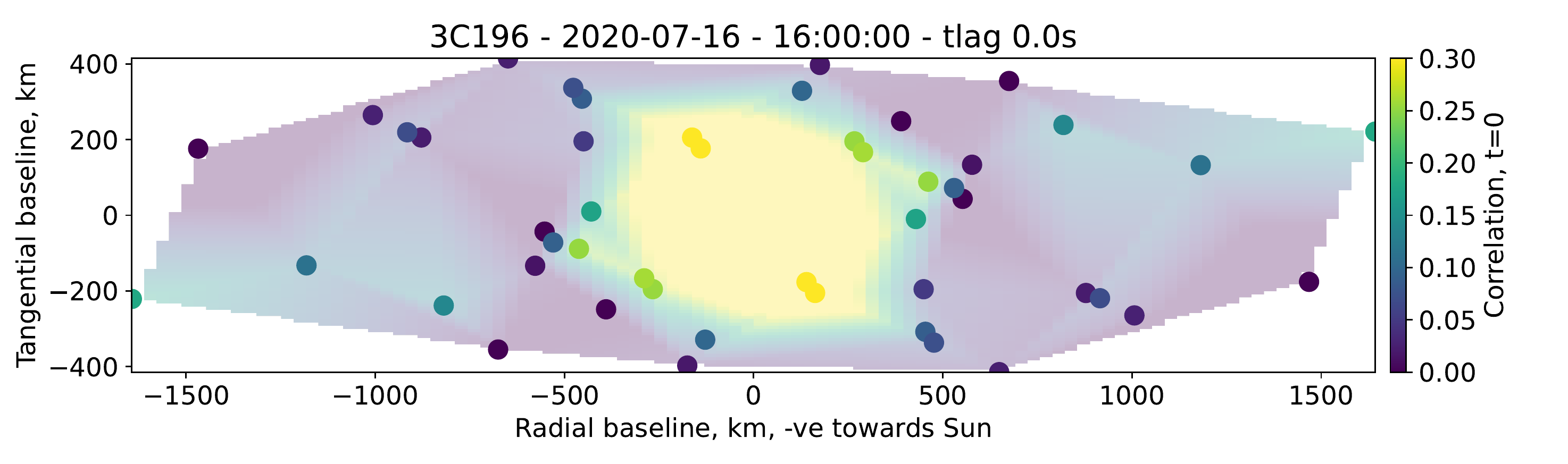}
    \caption{Spatial correlation image of zero time-lag values of the intensity cross-correlation functions, along with the individual data points used to form it.  The minimum and maximum cross-correlation values used in the colour scale are chosen to saturate the image such that the basic shape is emphasised.}
    \label{fig:zerolag}
\end{figure}

Example cross-correlation values at zero time lag for 16:00\,UT, prior to the occultation of 3C196 by the comet tail, are given in Figure \ref{fig:zerolag}.  The minimum and maximum cross-correlation values used in the colour scale are chosen to saturate the image such that the basic shape is emphasised, which is what is important here.  This shape is a result of the convolution of the structure of the radio source itself and that of the spatial spectrum of density irregularities giving rise to the IPS observed.  The individual values used to form the image are overplotted to illustrate how the contours were formed.  There is a dearth of data points with short baselines, resulting from the fact that data from several remote stations in the Netherlands were considered too poor for use, meaning that the central area of the image contains no data. However, additional structure is not expected in this region for a normal solar wind and the shape presented here mostly matches that of the solar wind given in \citet{Fallowsetal:2022a}.  

\begin{figure}
    \centering
    \includegraphics[width=\linewidth]{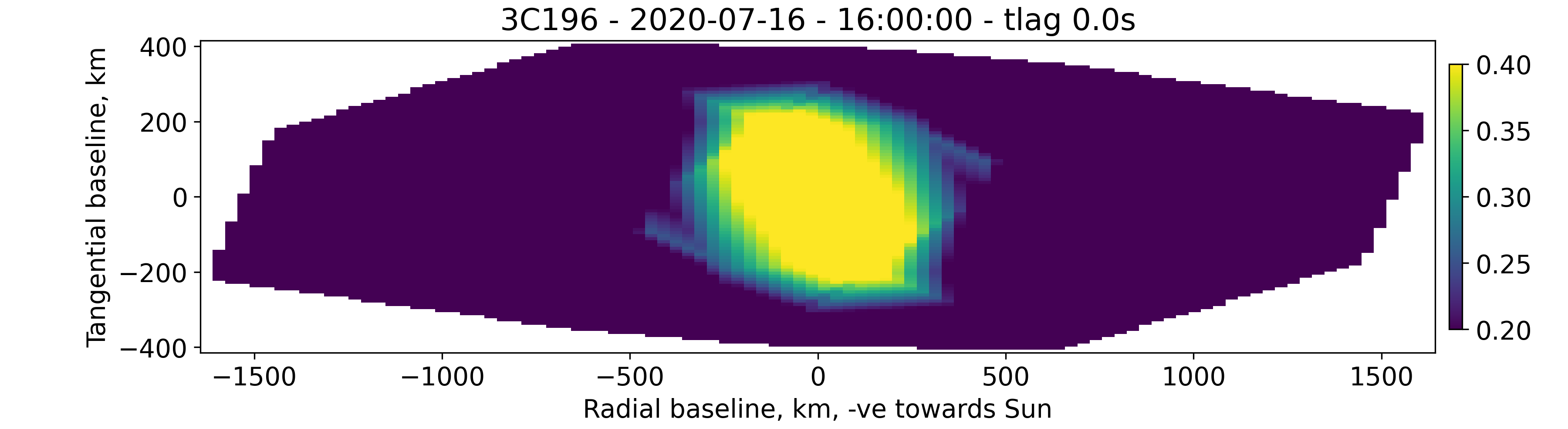}
    \includegraphics[width=\linewidth]{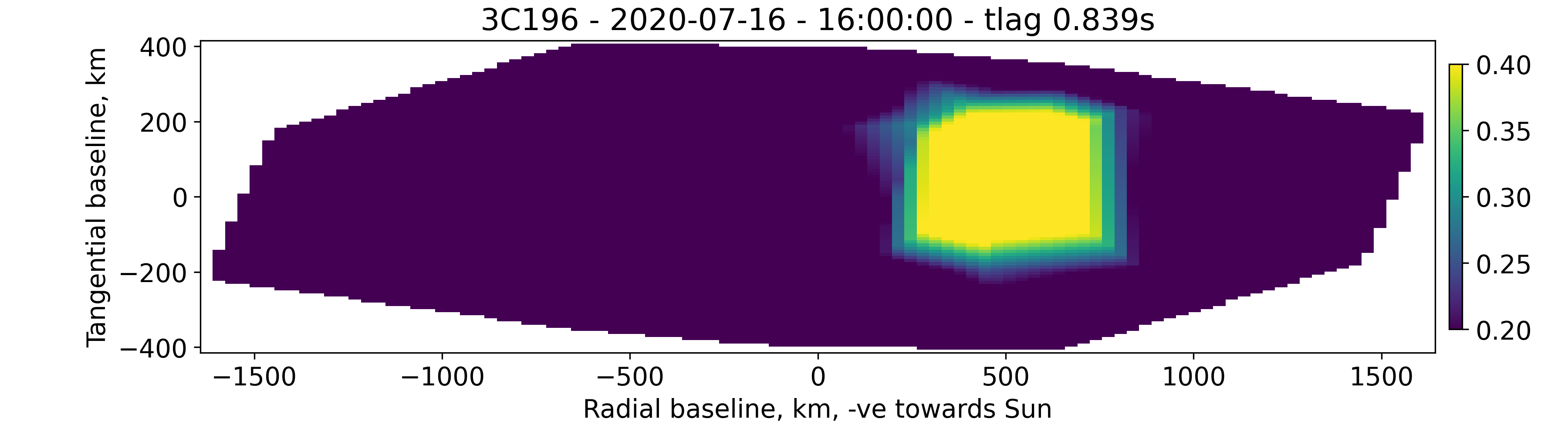}
    \includegraphics[width=\linewidth]{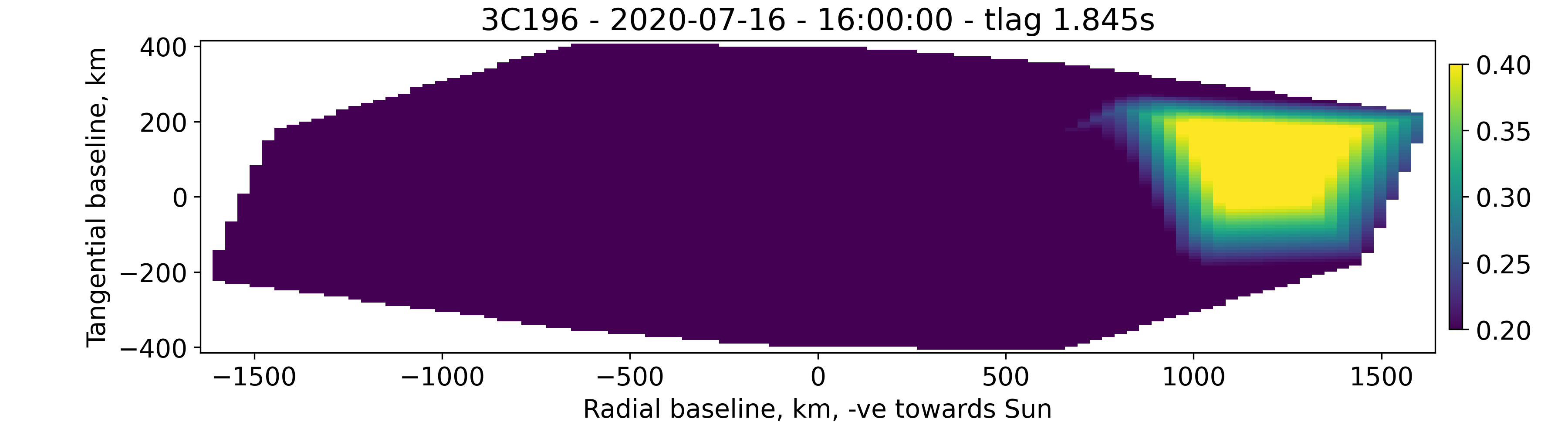}
    \caption{Spatial correlation images at time lags [top to bottom] 0\,s, 0.84\,s, and 1.85\,s for 16:00\,UT.}
    \label{fig:spatialtlags1600}
\end{figure}

The remainder of the CCF contains a great deal of additional information when viewed in spatial plots such as these, and by cycling through all the time lags, a movie of the spatial cross-correlation can be made that visualises not only structure but also motion.  Figure \ref{fig:spatialtlags1600} illustrates this for spatial correlation images at different time lags for 16:00\,UT, and shows how the structure moves with the solar wind across the field of view.  It should be noted that, although the non-zero time-lag images are less affected by radio source structure and are therefore more indicative of solar wind density structure, they are still strongly biased by the available data points used to form them.

\begin{figure*}
    \centering
    \includegraphics[width=0.49\linewidth]{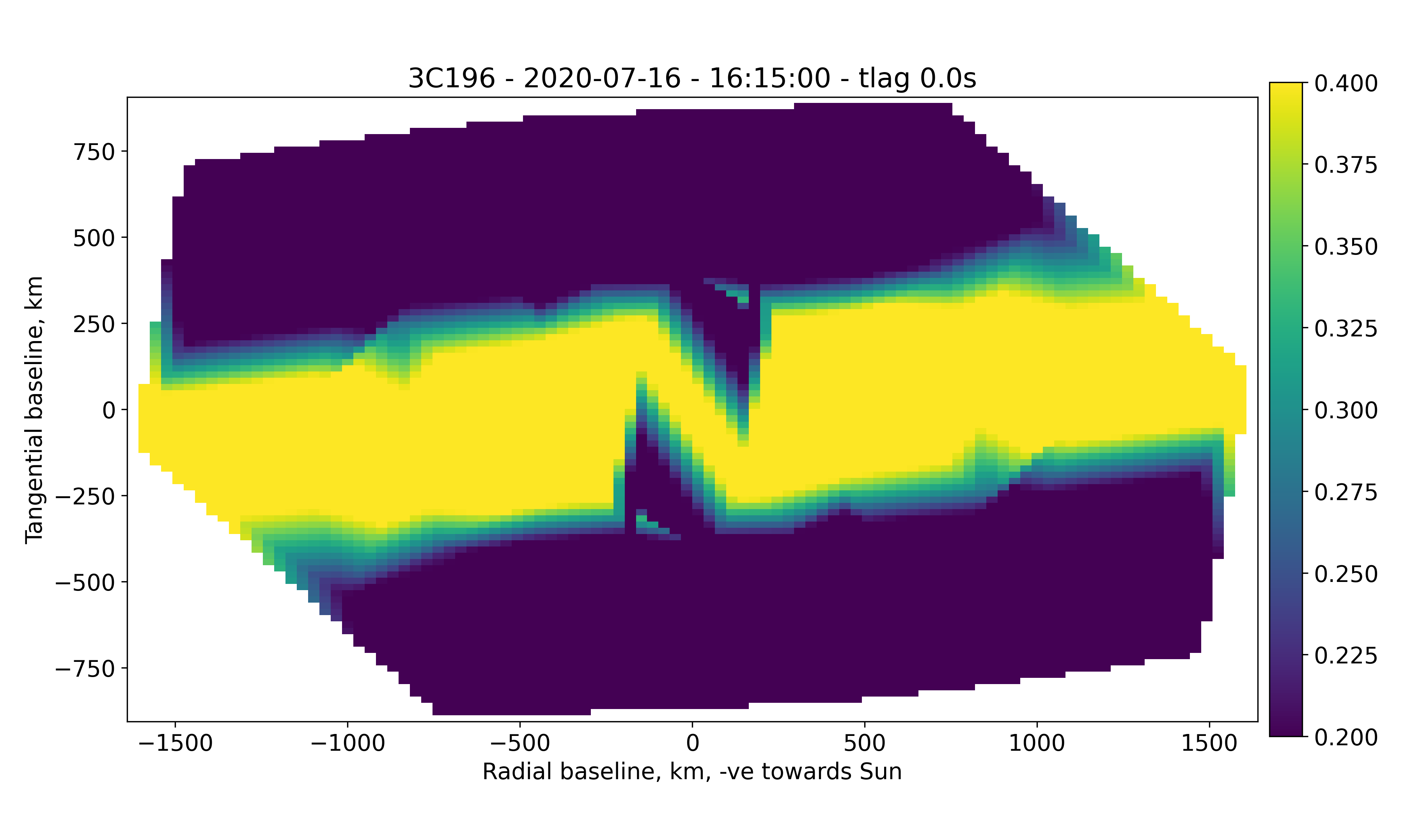}\includegraphics[width=0.49\linewidth]{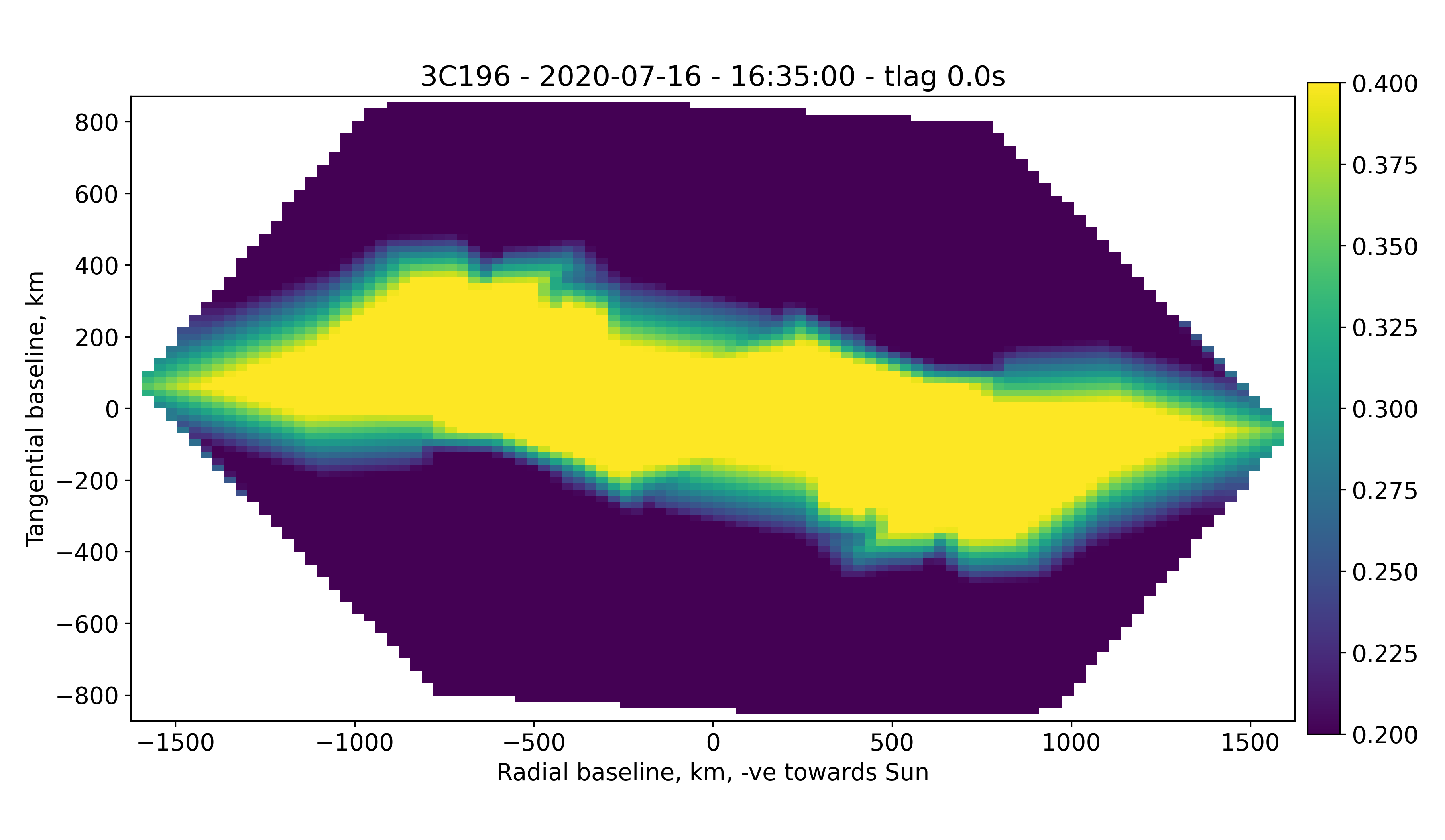}
    \includegraphics[width=0.49\linewidth]{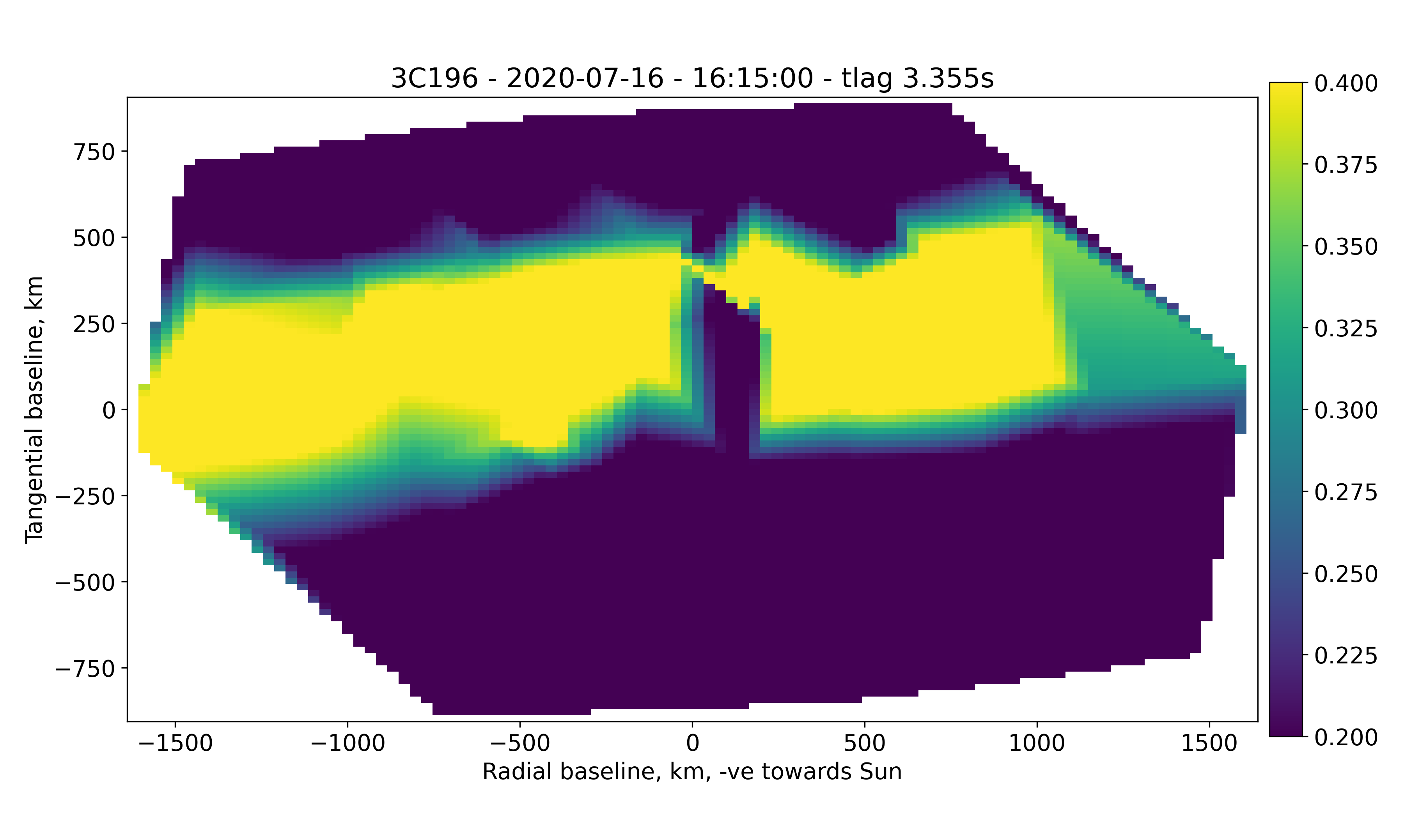}\includegraphics[width=0.49\linewidth]{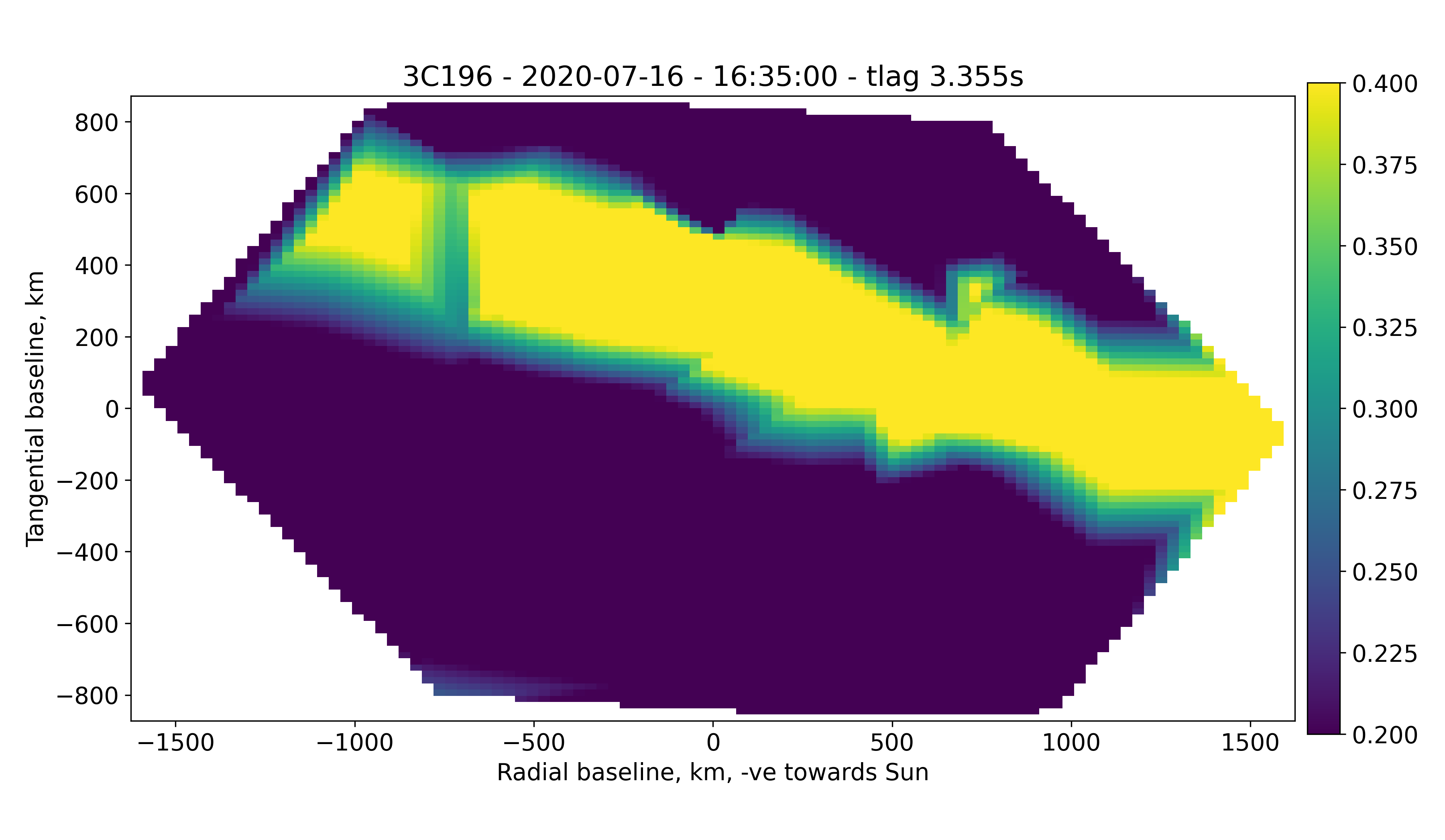}
    \includegraphics[width=0.49\linewidth]{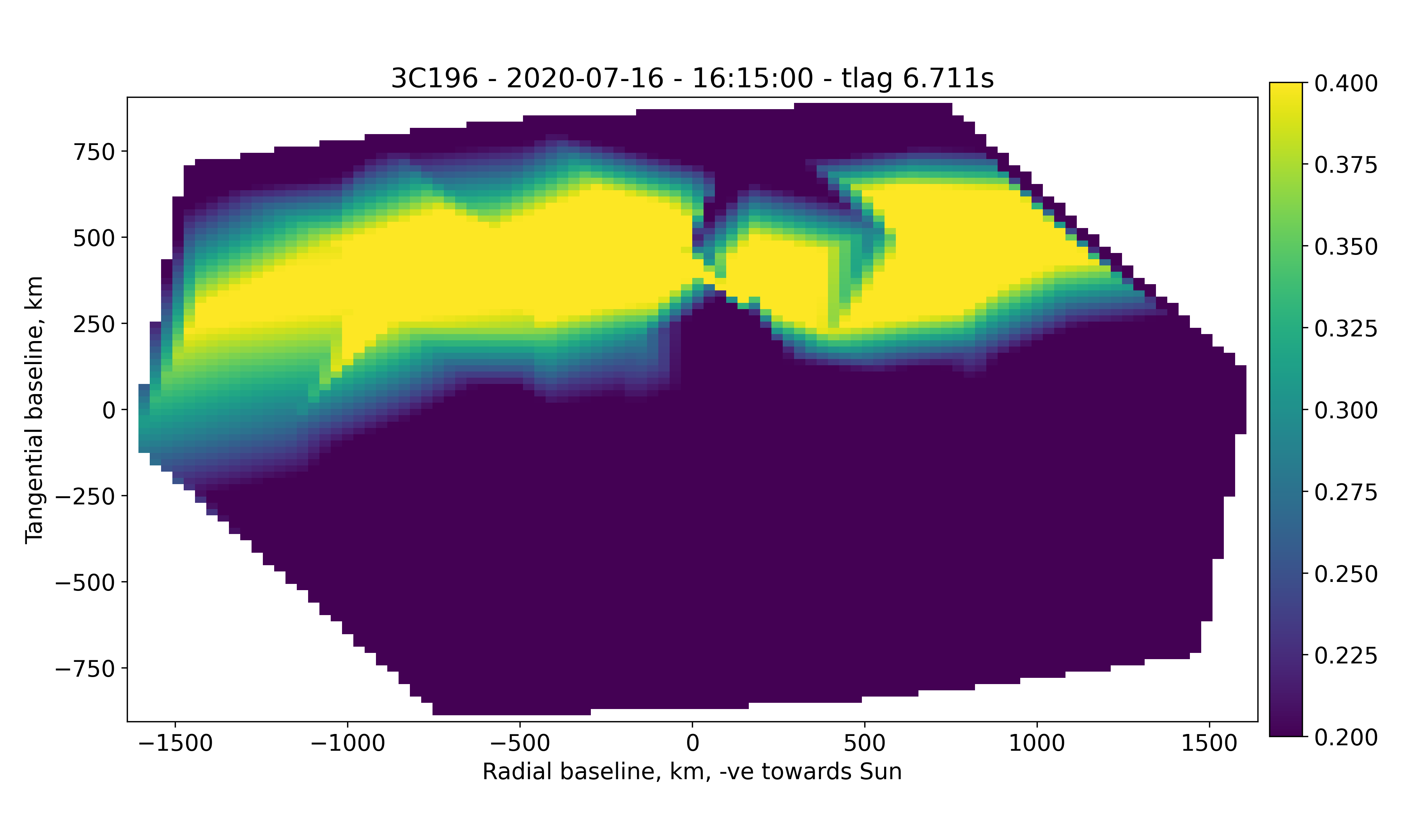}\includegraphics[width=0.49\linewidth]{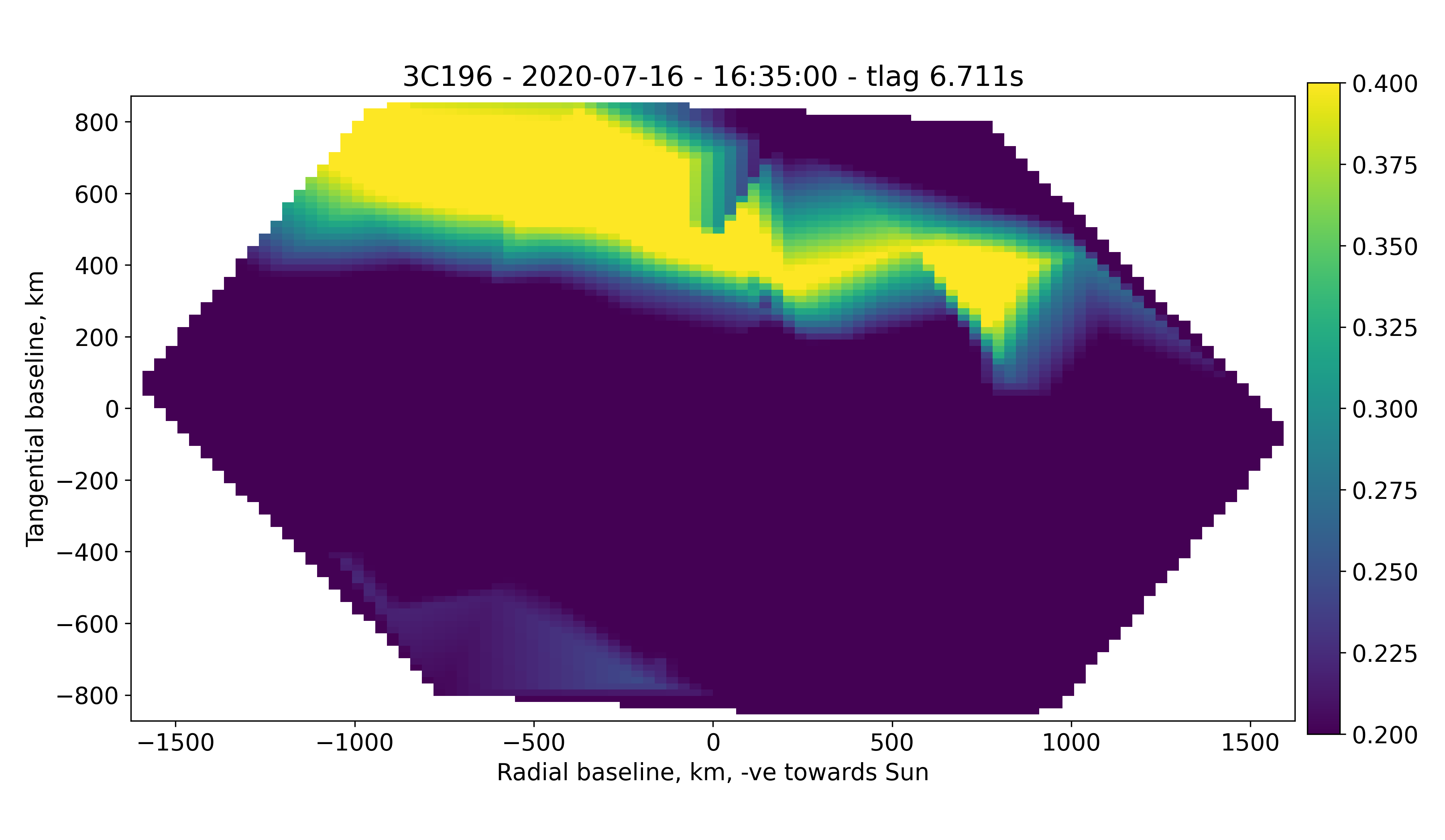}
    \caption{Spatial correlation images at time lags [top to bottom] 0\,s, 3.36\,s, and 6.71\,s for [left] 16:15\,UT and [right] 16:35\,UT. Movies of images at all available time lags are given in online material.}
    \label{fig:spatialtlags}
\end{figure*}

Figure \ref{fig:spatialtlags} presents several images at different time lags for 16:15\,UT and 16:35\,UT, times towards the start and end of the occultation, respectively.  These show a very different structure from that shown in Figure \ref{fig:spatialtlags1600} and motion within the structure cannot be clearly discerned in the images presented.  Gaps within the broad structure remain static and are therefore a result of the available data points.  However, motion is apparent in movies created from images at all time lags for these times and these are available in the online material.  The movies illustrate both the motion associated with the solar wind, similar to that shown in Figure \ref{fig:spatialtlags1600}, and a ---less clear but certainly appreciable--- slow-speed motion of material within the broader structure itself.

\section{Discussion}

The results presented show unequivocally that the enhanced scintillation seen here is a result of scattering through the plasma tail of the comet.  It appears only when the line of sight passes through the tail, is seen by all LOFAR stations ---thus discounting an ionospheric origin---, and is associated with a second velocity at a little over 100\,km\,s$^{-1}$, itself far slower than typically observed in the solar wind at these distances and too fast to be ionospheric.

The background solar wind stream is fast, at around 700\,km\,s$^{-1}$, which is expected given the high heliographic latitude of 66$^{\circ}$ of the point of closest approach to the Sun of the line of sight at the time.  There is some obvious variation during the period of enhanced scintillation which is not present at other times; the exact reason can only be postulated at this time, but perhaps it is a result of some buffeting between the solar wind and the comet tail.  The second velocity has a mean value of 123\,km\,s$^{-1}$, which is consistent with the value expected from within the comet tail close to the comet itself.  The location of the comet within the line of sight is known, which means that the foreshortening of the velocity can be corrected by division by cos(16.8) (the angle given in Figure \ref{fig:diagram}).  This correction is small, increasing the mean value to 128\,km\,s$^{-1}$.  The physical distance of the line of sight from the comet is estimated to be $\sim$860\,000\,km at closest approach.  This velocity therefore fits with what would be expected from spacecraft measurements covering a few different distances from their respective comets \citep[see e.g. a summary table presented in][]{Neugebaueretal:2007}, as well as the assumption of \cite{Ijuetal:2015}.

In the presence of weak scattering and under the assumption that radio-wave scintillation is produced from scattering through a single phase-changing screen, it can be shown that the amount of scintillation is in direct proportion to the strength of the small-scale variations in density, $\Delta N_{e}$.  This quantity is commonly assumed to be related to the density $N_{e}$ in observations of IPS \citep[e.g.][]{Jacksonetal:1998}, but the nature of this relationship is not exact for all circumstances.  \citet{Ijuetal:2015} assumed proportionality between $\Delta N_{e}$ and $N_{e}$ in their estimation of $N_{e}$ within the tail of comet ISON (C/2012 S1).  However, direct spacecraft measurements of density within the plasma tail \citep[e.g.][]{Bameetal:1986, Meyer-Vernetetal:1986} suggest a very sharp peak in density, and do not indicate any dip in the centre that could be consistent with the `hollow-tube' plasma distribution postulated by \citet{Birdetal:1984} and inferred from the $N_{e}$ results of \citet{Ijuetal:2015}.  The curved intensity structure in the dynamic spectra presented in Figure \ref{fig:dynspec} suggest refraction through the edges of a denser region of plasma and the application of a simple numerical scattering model \citep{Boydeetal:2022} to a single scattering screen with an assumed Gaussian density structure results in such a curved structure in a dynamic spectrum (Boyde, private communication).  The sharply peaked nature of the scintillation profile shown here therefore suggests a relation with strong turbulence along the tail boundary, but the dip in scintillation index between the peaks is unlikely to be associated with a dip in density, illustrating a case where the assumed proportionality between $\Delta N_{e}$ and $N_{e}$ breaks down.  Given that the velocity of the tail material remains almost constant and only detectable within the period of enhanced scintillation, and that the analysis of white-light measurements of a different comet plasma tail indicate a flow at solar wind speeds \citep[e.g.][]{Cloveretal:2010}, we postulate that a steep velocity shear exists between the tail boundary and the surrounding solar wind and that this is the likely cause of the strong turbulence.

The power spectra shown in Figure \ref{fig:pspec} also show the dual contributions from the solar wind and the plasma tail, with a low-frequency excess evident from all stations throughout the period of enhanced scintillation.  Under normal circumstances this excess could easily be confused with an ionospheric contribution (as illustrated by the Fresnel frequency calculations in Section \ref{subsec:pspec} and the theoretical treatment in \citet{Forteetal:2022}), illustrating the necessity for simultaneous data from multiple, widely spaced stations and/or observations of additional radio sources not occulted by the plasma tail to discount this possibility.  





The spatial correlation images shown in Figure \ref{fig:spatialtlags} provide further confirmation of the comet tail being the source of the enhanced scintillation.  The structure shown, which primarily reflects the small-scale density variations giving rise to scintillation, shows a highly elongated structure which broadly matches in direction the elongated tendrils of plasma seen in photographs such as that given in Figure \ref{fig:diagram}.  Furthermore, the spatial correlation structure can be seen to move across the field of view as the comet itself moves with respect to the LOFAR lines of sight.  In the movies, structure associated with the background solar wind can be seen propagating quickly out of the field of view in the same way as can be seen in Figure \ref{fig:spatialtlags1600}, while further substructure within the main elongated ribbon can be observed to propagate within it at the speed calculated for the cometary material.  These images illustrate that material within the plasma tail of the comet is not flowing exactly in a radial direction away from the Sun, but is slightly off-radial with the flow orientation being opposite on different sides of the tail.  The images at 16:35\,UT show a slightly greater orientation with respect to the radial direction than those at 16:15\,UT, which appears to reflect the orientations of plasma tendrils in the photograph.

\section{Conclusions}

This observation represents a unique combination of highly detailed data from a world-leading radio telescope and a fortuitous geometry whereby the radio source cuts an almost direct profile across the plasma tail of the comet.  The effects of scintillation due to the tail itself are unequivocally visible, enabling us to gain insight into the turbulence within the tail and the interaction between it and the surrounding solar wind.

The velocity of material within the plasma tail of the comet is found to be just over 100\,km\,$^{-1}$ at a distance of approximately 860\,000\,km downstream from the comet itself.  This velocity is detected simultaneously with the $\sim$700\,km\,$^{-1}$ of the background fast solar wind and remains constant throughout almost the entire period of enhanced scintillation. 

Scintillation indices show a twin-peak structure indicating strong turbulence along the tail boundaries.  When combined with the velocity information, this suggests that the turbulence is the result of a strong velocity shear between the comet tail and the surrounding solar wind.  The curved intensity structure in the dynamic spectra indicate refraction through the edges of a denser density structure, whereas the dip in scintillation indices between the twin peaks would typically suggest a density decrease, representing an occasion where the usual assumption of proportionality between $\Delta N_{e}$ and $N_{e}$ breaks down.

The spatial correlation images offer a unique view of the turbulent structure, and further confirm the association of the IPS observed with the comet tail.  The turbulent structure appears highly elongated in off-radial directions, which appear to match the tendrils of plasma seen in detailed photographs.

The duration of the enhanced scintillation in combination with the known distance to the comet leads to an estimate of the comet tail diameter of $\sim$100\,000\,km.  This is a tiny distance compared to the length of the line of sight through the inner heliosphere ($\sim$2\,AU), and yet the comet tail is both dense enough and turbulent enough to lead to an obvious change in the observed IPS, at least at this close pass of the comet to the line of sight.  A common assumption in the analysis of scintillation is that the scattering is due to a `thin screen' in the line of sight, with the line of sight integration accomplished through the addition of several such screens \citep[e.g.][]{Coles:1996}.  The scintillation from the narrow comet tail presented here can be pictured as being due to only a single screen in the line of sight, making this observation a natural laboratory with which to test scattering theory and associated assumptions applied to observations of radio scintillation covering any of the interstellar, interplanetary, and ionospheric regimes.

\begin{acknowledgements}
      This paper is based on data obtained with the International LOFAR Telescope (ILT) under project code DDT14\_001.  LOFAR \citep{vanHaarlemetal:2013} is the Low Frequency Array designed and constructed by ASTRON.  It has observing, data processing, and data storage facilities in several countries, that are owned by various parties (each with their own funding sources), and that are collectively operated by the ILT foundation under a joint scientific policy.  The ILT resources have benefitted from the following recent major funding sources: CNRS-INSU, Observatoire de Paris and Université d'Orléans, France; BMBF, MIWF-NRW, MPG, Germany; Science Foundation Ireland (SFI), Department of Business, Enterprise and Innovation (DBEI), Ireland; NWO, The Netherlands; The Science and Technology Facilities Council, UK; Ministry of Science and Higher Education, Poland.  Two of us (RAF and MMB) were partially supported by the LOFAR4SW project, funded by the European Community’s Horizon 2020 Programme H2020 INFRADEV-2017-1 under grant agreement 777442.  MMB was also supported in part by the STFC In-House Research grant to the Space Physics and Operations Division at UKRI STFC RAL Space.  BF was supported by the UK Natural Environment Research Council (Grant number NE/V002597/1).
\end{acknowledgements}

%
%

\bibliographystyle{aa}
\bibliography{comet_neowise}

\end{document}